\documentclass[conference]{IEEEtran}
\IEEEoverridecommandlockouts
\usepackage{cite}
\usepackage{amsmath,amssymb,amsfonts}
\usepackage{graphicx}
\usepackage{textcomp}
\usepackage{xcolor}
\def\BibTeX{{\rm B\kern-.05em{\sc i\kern-.025em b}\kern-.08em
    T\kern-.1667em\lower.7ex\hbox{E}\kern-.125emX}}

\usepackage{pgfplots}
\usepgfplotslibrary{fillbetween}
\usepgfplotslibrary{groupplots}
\pgfplotsset{compat=1.18}
\usetikzlibrary{positioning}
\usepackage{algorithm}
\usepackage[italicComments=false, indLines=false,commentColor=teal]{algpseudocodex}

\usepackage[caption=false,font=footnotesize]{subfig}
\usepackage{orcidlink}
\hypersetup{hidelinks}

\begin{document}

\title{Mode-Tensorized Canonical Polyadic Decomposition for MIMO Channel Estimation\\
}

\author{
\IEEEauthorblockN{Alexander Blagodarnyi \orcidlink{0000-0002-8238-9464}, Alexander Sherstobitov \orcidlink{0000-0002-5489-0155}, Vladimir Lyashev \orcidlink{0000-0002-6993-1589}}
\IEEEauthorblockA{
The Dept. of Multimedia Technology and Telecommunication, Moscow Institute of Physics and Technology\\
Moscow, Russia\\
blagodarnyi.as@phystech.edu, sherstobitov@rtt-lab.ru, lyashev.va@mipt.ru
}
}
\maketitle
\begin{abstract}
This paper proposes a channel estimation method for Multiple-Input Multiple-Output (MIMO) systems based on Canonical Polyadic (CP) decomposition applied to a mode-factorized tensor representation of the channel. The proposed approach reshapes the original low-order channel tensor into a higher-order tensor by factorizing its modes into multiple virtual modes, thereby introducing additional dimensions. By exploiting the sparse structure of MIMO channels and the plane-wave propagation model in the far-field regime, the proposed mode tensorization enhances the separability of individual propagation paths. It is shown that increasing the number of tensor modes improves component separation and provides inherent denoising effects. Building on these properties, a mode-tensorized CP decomposition (MTCPD) algorithm is developed. In addition, a metric for analyzing the virtual factors obtained from MTCPD is proposed, enabling estimation of the canonical rank and selection of the most informative components contributing to overall system performance. Numerical results demonstrate that the proposed method improves channel estimation accuracy compared to conventional tensor-based approaches, particularly under low signal-to-noise ratio conditions.
\end{abstract}


\section{Introduction}
Multiple-input multiple-output orthogonal frequency-division multiplexing (MIMO-OFDM) is a key technology in modern wireless communication systems, including LTE and 5G New Radio (NR), that exploits spatial diversity to improve system throughput and coverage. Reliable operation of such systems critically depends on accurate channel estimation (CE), which enables the acquisition of channel state information (CSI). The CSI is typically obtained from uplink pilot transmissions and is inherently corrupted by additive white Gaussian noise (AWGN). Accurate CSI is essential for subsequent tasks such as beamforming and downlink precoding \cite{MIMO_book}.

Channel estimation in MIMO systems is a challenging problem due to the high dimensionality of the channel. The MIMO channel can be naturally represented as a multidimensional tensor, whose size rapidly increases with the number of antennas and system bandwidth, leading to a significant growth in the number of parameters to be estimated, stored, and processed \cite{T_CE1}.

To address this problem, tensor decomposition techniques have emerged as a promising framework for MIMO channel estimation in noisy environments \cite{T_CE1, T_CE_MA_CPD}. In particular, Canonical Polyadic Decomposition (CPD) provides a compact representation of the channel tensor as a sum of rank-one components for mmWave and terahertz systems \cite{T_CE2, T_CE_XL_CPD, T_CE_CPD_ISAC}. This representation is well aligned with the physical propagation model, where the channel consists of a limited number of dominant paths. As a result, MIMO channel tensors often exhibit a low-rank structure that can be exploited for efficient estimation and denoising.

However, computing CPD is a non-convex optimization problem typically solved using iterative algorithms such as Alternating Least Squares (ALS) and its variants. These methods require several design choices, including decomposition rank selection, factor initialization, and stopping criteria \cite{Ballard_Kolda_2025, Tensor_book_Lu}. In particular, the rank is usually assumed to be known in advance, which is rarely true in practice. 

Incorrect rank selection or poor CP-ALS initialization may lead to performance degradation due to noise overestimation and convergence to poor local minima \cite{CPD_init, CPD_init2}.

In contrast to mmWave and terahertz communications, sub-6 GHz MIMO channels are typically less sparse due to lower angular and delay resolution \cite{sub6g_sparsity}. This creates additional challenges for CPD-based channel estimation, since the estimated factor vectors may represent groups of closely spaced paths rather than individual propagation components.

In this paper, we address these challenges by introducing a mode-factorized tensor representation of the MIMO channel, where each tensor mode is decomposed into multiple sub-modes, effectively increasing the tensor order. This transformation introduces additional virtual dimensions that improve the separability of channel components and enhance CPD robustness to noise. Building on this representation, we propose a sequential rank-1 CP decomposition algorithm with adaptive rank determination based on analysis of the extracted factors.

\textit{Notations:} Scalars, vectors, matrices, and tensors are denoted by lowercase and uppercase letters (e.g., $x, X$), bold lowercase letters (e.g., $\mathbf{x}$), bold uppercase letters (e.g., $\mathbf{X}$), and calligraphic uppercase letters (e.g., $\mathcal{X}$), respectively, unless stated otherwise. Bold italic symbols (e.g., $\boldsymbol{X}$) denote tuples of dimensions or indices. The operators $(\cdot)^{\mathrm{T}}$, $(\cdot)^{\mathrm{H}}$, and $\|\cdot\|_F$ denote transpose, Hermitian transpose, and the Frobenius norm. The symbols $\circ$ and $\otimes$ denote the outer product and Kronecker product, respectively. Noisy and estimated quantities are denoted by $\tilde{(\cdot)}$ and $\hat{(\cdot)}$, respectively.

\section{System Model}
\subsection{Channel Model}

\begin{figure}
    \centering
    \includegraphics[width=0.97\linewidth,trim={0 0 3.5mm 1mm}, clip]{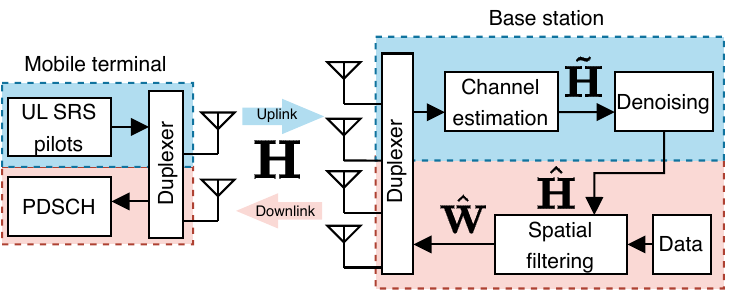}
    \caption{General system model illustrating downlink precoder design based on uplink channel estimation.}
    \label{fig:system_model}
\end{figure}

Consider a MIMO-OFDM system where the base station (BS) is equipped with a uniform rectangular array (URA) consisting of $X$ horizontal and $Y$ vertical antenna elements, resulting in a total of $M = XY$ antennas. The user equipment (UE) is equipped with $N$ antennas. We consider uplink transmission in a 5G New Radio (NR) time-division duplex (TDD) system, where the BS receives sounding reference signals (SRS) transmitted by the UE (Fig. \ref{fig:system_model}).

Due to multipath propagation, the channel is modeled as a superposition of $L$ propagation paths. Under the far-field assumption, each path can be approximated as a plane wave falling on the antenna array. The channel matrix $\mathbf{H}(t) \in \mathbb{C}^{M \times N}$ at time $t$ is expressed as
\begin{equation}
\mathbf{H}(t) =
\sum_{l=1}^{L}
s_l \,
\mathbf{a}_{\mathrm{BS}}(\alpha_l, \beta_l)
\mathbf{a}_{\mathrm{UE}}(\gamma_l)^{\mathrm{T}}
\delta(t - \tau_l),
\end{equation}
where $s_l \in \mathbb{C}$ denotes the complex gain of the $l$-th path and $\tau_l$ is the corresponding path delay.
The parameters $\alpha_l, \beta_l, \gamma_l \in \big[-\frac{1}{2}, \frac{1}{2}\big)$ denote normalized spatial frequencies corresponding to projections of the propagation direction onto the antenna array axes. This parametrization avoids explicit dependence on a particular angular coordinate system.

The UE array response vector $\mathbf{a}_{\mathrm{UE}}(\gamma_l) \in \mathbb{C}^{N}$ is defined as
\begin{equation}
\mathbf{a}_{\mathrm{UE}}(\gamma_l)=\left[1, e^{j2\pi \gamma_l}, \ldots, e^{j2\pi (N-1)\gamma_l} \right]^{\mathrm{T}}.
\end{equation}

The BS array response vector $\mathbf{a}_{\mathrm{BS}}(\alpha_l, \beta_l) \in \mathbb{C}^{M}$ admits a separable structure due to the URA geometry:
\begin{equation}
\mathbf{a}_{\mathrm{BS}}(\alpha_l, \beta_l)
=
\mathbf{a}_X(\alpha_l) \otimes \mathbf{a}_Y(\beta_l),
\end{equation}
where
\begin{equation}
\mathbf{a}_X(\alpha_l) =
\left[1, e^{j2\pi \alpha_l}, \ldots, e^{j2\pi (X-1)\alpha_l} \right]^{\mathrm{T}},
\end{equation}
\begin{equation}
\mathbf{a}_Y(\beta_l)
=
\left[1, e^{j2\pi \beta_l}, \ldots, e^{j2\pi (Y-1)\beta_l} \right]^{\mathrm{T}}.
\end{equation}

Assuming OFDM transmission with total bandwidth $B$ divided into $K$ subcarriers with spacing $\Delta f = B/K$, the channel is represented in the frequency domain. For a fixed OFDM symbol, the frequency response associated with delay $\tau_l$ is given by
\begin{equation}
\mathbf{g}(\tau_l)
=
\left[1, e^{j2\pi \tau_l}, \ldots, e^{j2\pi (K-1)\tau_l} \right]^{\mathrm{T}},
\end{equation}
where $\tau_l \in \big[-\frac{1}{2}, \frac{1}{2}\big)$ denotes the normalized delay.

Stacking the channel matrices across all subcarriers results in a third-order channel tensor $\mathcal{H}\in\mathbb{C}^{M \times N \times K}$, which admits the following low-rank decomposition:
\begin{equation}
\mathcal{H}
=
\sum_{l=1}^{L}
s_l
\,
\mathbf{a}_{\mathrm{BS}}(\alpha_l, \beta_l)
\circ
\mathbf{a}_{\mathrm{UE}}(\gamma_l)
\circ
\mathbf{g}(\tau_l),
\end{equation}

It can be observed that all factor vectors $\mathbf{a}_{\mathrm{BS}}$, $\mathbf{a}_{\mathrm{UE}}$, and $\mathbf{g}$ exhibit a Vandermonde (steering vector) structure parameterized by spatial or delay-domain frequencies \cite{Vandermonde_channel}.

In the presence of additive white Gaussian noise (AWGN), the observed tensor is
\begin{equation}
\tilde{\mathcal{H}} = \mathcal{H} + \mathcal{Z},
\end{equation}
where
$\mathcal{Z} \in \mathbb{C}^{M \times N \times K}$
is a complex Gaussian noise tensor with i.i.d. entries distributed as
$\mathcal{CN}(0,\sigma_{\mathrm{UL}}^2)$.

\subsection{Problem Formulation}

Given the noisy observation tensor
$\tilde{\mathcal{H}}$,
the objective is to estimate the underlying low-rank channel tensor
$\mathcal{H}$
and its constituent components corresponding to individual propagation paths. This problem can be formulated as tensor denoising and decomposition, where the goal is to recover a structured low-rank approximation
$\hat{\mathcal{H}}$
that preserves the physical interpretability of the channel while suppressing noise.

\section{Canonical Polyadic Decomposition for Channel Estimation}
In practical systems, the number of UE antennas $N$ is typically small, and the UE array structure may be unknown. Therefore, consider that channel denoising is performed independently for each UE antenna. Let
\begin{equation}
\tilde{\mathcal{H}} = [\tilde{\mathbf{H}}_1,\ldots,\tilde{\mathbf{H}}_N], 
\qquad
\tilde{\mathbf{H}}_n \in \mathbb{C}^{M \times K},
\label{eq:slice}
\end{equation}
where $\tilde{\mathbf{H}}_n$ denotes the observed channel corresponding to the $n$-th UE antenna.

Since the BS array geometry is known, the antenna dimension $M$ is reshaped according to the URA structure using the operator $\mathcal{R}(\cdot)$:
\begin{equation}
\tilde{\mathcal{X}}_n = 
\mathcal{R}(\tilde{\mathbf{H}}_n)
\in \mathbb{C}^{X \times Y \times K},
\qquad
M = XY.
\label{eq:reshape}
\end{equation}

The canonical polyadic (CP) decomposition represents $\tilde{\mathcal{X}}_n$ as a sum of $R$ rank-1 tensors \cite{Ballard_Kolda_2025}:
\begin{equation}
\tilde{\mathcal{X}}_n \approx
\sum_{r=1}^{R} \mathbf{u}_{n,r}^{(X)} \circ \mathbf{u}_{n,r}^{(Y)} \circ \mathbf{u}_{n,r}^{(K)},
\end{equation}
where
$\mathbf{u}_{n,r}^{(X)} \in \mathbb{C}^{X}$, $\mathbf{u}_{n,r}^{(Y)} \in \mathbb{C}^{Y}$, and $\mathbf{u}_{n,r}^{(K)} \in \mathbb{C}^{K}$ denote the canonical factor vectors associated with the $r$-th component.

Accordingly, channel estimation can be formulated as the problem of finding a low-rank CP approximation of
$\tilde{\mathcal{X}}_n$:
\begin{equation}
\underset{ \mathbf{u}_{n,r}^{(X)}, \mathbf{u}_{n,r}^{(Y)}, \mathbf{u}_{n,r}^{(K)}}{\mathrm{minimize}}
\Big\|
\tilde{\mathcal{X}}_n - \sum_{r=1}^{R} \mathbf{u}_{n,r}^{(X)} \circ \mathbf{u}_{n,r}^{(Y)} \circ \mathbf{u}_{n,r}^{(K)} \Big\|_F^2,
\end{equation}
solving leads to estimated factor vectors $\hat{\mathbf{u}}_{n,r}^{(X)}, \hat{\mathbf{u}}_{n,r}^{(Y)}, \hat{\mathbf{u}}_{n,r}^{(K)}$. 
Finally, the estimated channel tensor $\hat{\mathcal{X}}_n$ is expressed as:
\begin{equation}
\hat{\mathcal{X}}_n = \sum_{r=1}^R \hat{\mathcal{X}}_{n,r} =
\sum_{r=1}^R 
\hat{\mathbf{u}}_{n,r}^{(X)} \circ 
\hat{\mathbf{u}}_{n,r}^{(Y)} \circ 
\hat{\mathbf{u}}_{n,r}^{(K)}.
\label{eq:ch_cp_est}
\end{equation}
This nonconvex optimization problem is typically solved using iterative algorithms such as alternating least squares (ALS).

From the physical channel model, each tensor $\tilde{\mathcal{X}}_n$ is composed of multiple rank-1 components corresponding to individual propagation paths. However, the true number of paths is generally unknown, and closely spaced paths may not be resolvable due to limited spatial and frequency resolution. As a result, the effective CP rank $R$ may differ from the actual number of physical paths $L$, depending on the propagation conditions and system parameters.

Moreover, CP decomposition of $\tilde{\mathcal{X}}_n$ does not guarantee that each estimated rank-1 component corresponds to a single propagation path. In practice, one CP component may represent several different paths, which limits the ability of conventional CPD to fully exploit the underlying Vandermonde structure of the channel.

These limitations motivate the proposed mode-tensorized approach presented in the next section.
\section{Mode-Tensorized Canonical Polyadic Decomposition}

We first introduce the proposed tensorization concept using an idealized steering vector, and then extend it to the general channel tensor.

\begin{figure*}[!t]
    \centering
    \includegraphics[width=\textwidth,trim={6mm 1mm 5mm 6mm}, clip]{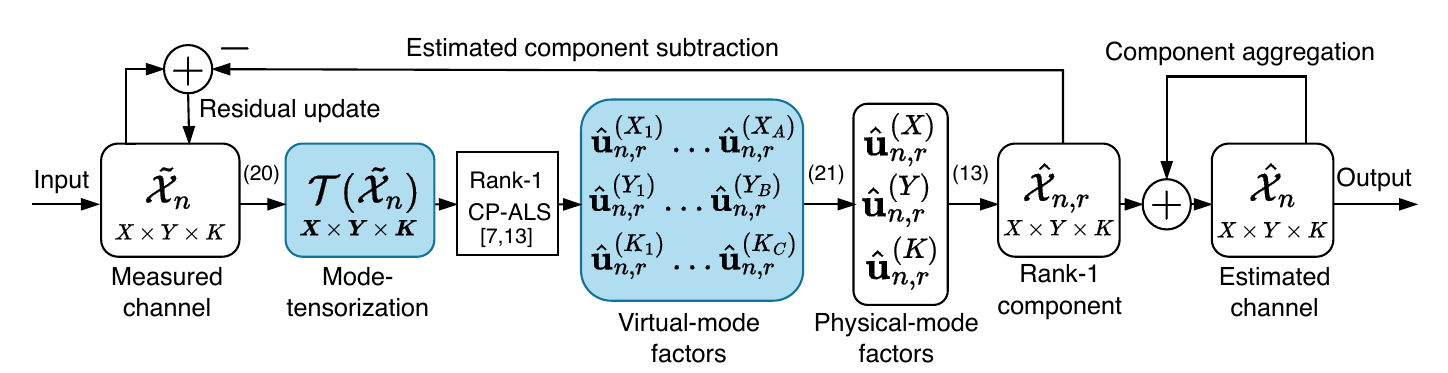}
    \caption{Block diagram of the proposed MTCPD-based rank-1 extraction. The blocks highlighted in blue correspond to the additional operations introduced by MTCPD; removing these blocks yields the conventional CPD-based pipeline.}
    \label{fig:mtcpd_scheme}
\end{figure*}

\subsection{Tensorization of Steering Vectors}

Consider a steering vector of length $K$ parameterized by $\alpha$:
\begin{equation}
\mathbf{v}^{(K)}(\alpha) = 
\left[1, e^{j2\pi \alpha}, \ldots, e^{j2\pi (K-1)\alpha}\right]^{\mathrm{T}}.
\end{equation}

Assume that $K$ admits a factorization into $C$ integers,
\begin{equation}
K = \prod_{c=1}^{C} K_c, \quad \boldsymbol{K} = (K_1, \ldots, K_C).
\end{equation}

We define a \textit{mode-tensorization} operator $\mathcal{T}_{\boldsymbol{K}}$ that reshapes $\mathbf{v}^{(K)}(\alpha)$ into a $C$-th order tensor \cite{2025_blagodarnyi}
\begin{equation}
\mathcal{V} = \mathcal{T}_{\boldsymbol{K}}(\mathbf{v}^{(K)}) 
\in \mathbb{C}^{K_1 \times \cdots \times K_C}.
\end{equation}

Due to the exponential (Vandermonde) structure of $\mathbf{v}^{(K)}(\alpha)$, the tensorized representation $\mathcal{V}$ admits an exact rank-one decomposition:
\begin{equation}
\mathcal{V} = 
\mathbf{v}^{(K_1)}(\alpha_1) \circ 
\mathbf{v}^{(K_2)}(\alpha_2) \circ \cdots \circ 
\mathbf{v}^{(K_C)}(\alpha_C),
\end{equation}
where $\alpha_c$ are scaled versions of $\alpha$.

Equivalently, the original steering vector can be represented as a Kronecker product of lower-dimensional steering vectors \cite{2025_blagodarnyi}:
\begin{equation}
\mathbf{v}^{(K)}(\alpha) = 
\bigotimes_{c=1}^{C} \mathbf{v}^{(K_c)}(\alpha_c).
\label{eq:kron_fact}
\end{equation}

This shows that tensorization transforms a steering vector into a structured rank-one tensor whose factors preserve the Vandermonde form. This property serves as the foundation of the proposed approach.

\subsection{Extension to Channel Tensor}

The above derivation holds exactly for ideal steering vectors. However, in practical propagation environments, channel responses deviate from this model due to multipath propagation with closely spaced subpaths and the presence of noise. As a result, the tensorized representation is no longer strictly rank-one, and CP decomposition does not guarantee recovery of ideal Vandermonde factors.

Nevertheless, the underlying exponential structure remains approximately valid. Tensorization redistributes this structure across multiple modes, which improves the separability of components in CP decomposition.

Let us introduce factorizations of each mode of the tensor $\tilde{\mathcal{X}}_n \in \mathbb{C}^{X \times Y \times K}$:
\begin{equation}
X = \prod_{a=1}^{A} X_a, \quad
Y = \prod_{b=1}^{B} Y_b, \quad
K = \prod_{c=1}^{C} K_c,
\end{equation}
with $\boldsymbol{X} = (X_1,\ldots, X_A)$, $\boldsymbol{Y} = (Y_1, \ldots, Y_B)$, and $\boldsymbol{K}=(K_1, \ldots, K_C)$.
Applying the tensorization operator yields a higher-order tensor:
\begin{equation}
\mathcal{T}(\tilde{\mathcal{X}}_n) \in 
\mathbb{C}^{X_1 \times \cdots \times X_A \times 
             Y_1 \times \cdots \times Y_B \times 
             K_1 \times \cdots \times K_C}.
             \label{eq:tensorization}
\end{equation}
For simplicity, we denote this space as $\mathbb{C}^{\boldsymbol{X}\times\boldsymbol{Y}\times\boldsymbol{K}}$.
A CP decomposition of $\mathcal{T}(\tilde{\mathcal{X}}_n)$ produces a set of virtual-mode factors
$\hat{\mathbf{u}}_{n,r}^{(X_a)}$, $\hat{\mathbf{u}}_{n,r}^{(Y_b)}$, and $\hat{\mathbf{u}}_{n,r}^{(K_c)}$. These factors are combined to reconstruct the physical-mode factors using the Kronecker structure in \eqref{eq:kron_fact}:
\begin{equation}
\hat{\mathbf{u}}_{n,r}^{(X)} = \bigotimes_{a=1}^{A} \hat{\mathbf{u}}_{n,r}^{(X_a)}, \quad
\hat{\mathbf{u}}_{n,r}^{(Y)} = \bigotimes_{b=1}^{B} \hat{\mathbf{u}}_{n,r}^{(Y_b)}, \quad
\hat{\mathbf{u}}_{n,r}^{(K)} = \bigotimes_{c=1}^{C} \hat{\mathbf{u}}_{n,r}^{(K_c)}.
\label{eq:virt2phys}
\end{equation}
Finally, the estimated channel $\hat{\mathcal{X}}_n$ can be obtained from physical mode factors $\hat{\mathbf{u}}_{n,r}^{(X)}, \hat{\mathbf{u}}_{n,r}^{(Y)}, \hat{\mathbf{u}}_{n,r}^{(K)}$ using (\ref{eq:ch_cp_est}).

Analysis of the proposed tensorization introduces an implicit regularization effect. In conventional CPD applied directly to $\tilde{\mathcal{X}}_n$, the number of parameters scales as $R(X+Y+K)$ \cite{Ballard_Kolda_2025}. In contrast, after tensorization, the number of estimated parameters becomes $R\left(\sum_{a=1}^A X_a + \sum_{b=1}^B Y_b + \sum_{c=1}^C K_c\right)$.

This reduction becomes more pronounced as the tensorization order increases. In the extreme case where all sub-dimensions are equal to 2 (i.e., $X_a=Y_b=K_c=2$), the number of parameters scales logarithmically as $R(\log_2(XYK)+1)$ after normalization of factor vectors.

This implicit regularization suppresses noise while preserving the underlying Vandermonde structure of the channel.

\section{MTCPD and CPD Algorithms}
This section describes practical modifications of the CPD-ALS algorithm that account for the structural properties of MIMO channel responses. These modifications are applied to both the conventional CPD and the proposed MTCPD.

Due to the nonconvex nature of the CPD-ALS optimization problem, the algorithm is sensitive to factor initialization \cite{ CPD_init, CPD_init2}. Standard random initialization may lead to slow convergence and poor local minima, particularly at low signal-to-noise ratio (SNR), where noise introduces multiple spurious solutions.

To improve robustness, the initial factors are chosen from columns of the discrete Fourier transform (DFT) matrix corresponding to dominant frequency components. Such initialization promotes alignment with the underlying signal subspace and accelerates convergence of the ALS iterations. \cite{2025_blagodarnyi}

Another modification addresses the spectral leakage effect, where energy from a single propagation path may spread across neighboring components, leading to redundant factor estimates. To mitigate this issue, a rank-by-rank extraction strategy is employed, in which rank-one components are iteratively estimated and removed from the residual tensor \cite{2025_blagodarnyi}. 

Both the proposed MTCPD and the conventional CPD employ a rank-by-rank extraction strategy with initialization based on discrete Fourier transform (DFT) vectors. The Block diagram of the researched pipeline is shown in Figure \ref{fig:mtcpd_scheme}, the description is presented in Algorithm \ref{alg:mtcpd}.

\begin{algorithm}[t]
\caption{MTCPD channel estimation for one UE slice}
\label{alg:mtcpd}
\begin{algorithmic}[1]
\Require Reshaped noisy tensor $\tilde{\mathcal{X}}_n$,
tensorization parameters,
maximum rank $R$
\Ensure Estimated tensor $\hat{\mathcal{X}}_n$
\For{$r=1,\ldots,R$}
    \State Apply to $\tilde{\mathcal{X}}_n$ mode tensorization as in (\ref{eq:tensorization})
    \State Initialize factors with dominant DFT vectors as in \cite{2025_blagodarnyi}
    \State Extract rank-1 component using CP-ALS \cite{Ballard_Kolda_2025} 
    \State Reconstruct physical-mode factors as in (\ref{eq:virt2phys})
    \State Form rank-1 component $\hat{\mathcal{X}}_{n,r}$
    \State Update residual $\tilde{\mathcal{X}}_n \gets \tilde{\mathcal{X}}_n-\hat{\mathcal{X}}_{n,r}$
\EndFor
\State Reconstruct
$\hat{\mathcal{X}}_n=\sum_{r=1}^{R} \hat{\mathcal{X}}_{n,r}$

\end{algorithmic}
\end{algorithm}

\section{Experimental modeling}

\begin{table}[t]
\centering
\caption{Simulation Parameters}
\label{tab:sim_params}
\begin{tabular}{|l|c|c|}
\hline
\textbf{Parameter} & \multicolumn{2}{c|}{\textbf{Value}} \\ \hline

Scenario & \multicolumn{2}{c|}{3GPP TR 38.901 UMa NLOS} \\
Carrier frequency & \multicolumn{2}{c|}{3.6 GHz} \\
Bandwidth / Subcarriers ($K$) & \multicolumn{2}{c|}{$100$ MHz / $512$} \\
Realizations $|\mathcal{D}|$ & \multicolumn{2}{c|}{1200} \\
BS array ($X \times Y$) & \multicolumn{2}{c|}{$8 \times 8$ ($M=64$)} \\
UE array ($N$) & \multicolumn{2}{c|}{2} \\ \hline 

\multicolumn{3}{|c|}{\textit{Algorithm Parameters}} \\ \hline
 & \textbf{CPD} & \textbf{MTCPD} \\ \hline
Mode decomposition $(A,B,C)$ & $(1,1,1)$ & $(3,3,9)$ \\
Sub-dimensions $(X_a, Y_b, K_c)$ & $(8,8,512)$ & $(2,2,2)$ $\forall a,b,c$ \\ \hline
Max rank ($R$) & \multicolumn{2}{c|}{40} \\
Max ALS iterations & \multicolumn{2}{c|}{1000} \\
Tolerance & \multicolumn{2}{c|}{$10^{-6}$} \\ \hline
\end{tabular}
\end{table}

The channel data were generated using the QuaDRiGa simulator under the 3GPP TR 38.901 UMa NLOS scenario \cite{QDG}. The ideal channel tensor was normalized to $\|\mathcal{H}\|_F=1$, while the norm of the noise tensor $\mathcal{Z}$ depends on the fixed SNR value. Described channel denoising algorithms were validated in the range of $[-24, 4]$ dB Uplink SNR values. The main simulation parameters are summarized in Table~\ref{tab:sim_params}.

\subsection{Dataset-Averaged Rank Selection}

As a reference analysis, we consider a dataset-level rank selection strategy under the assumption that the ideal channel tensor $\mathcal{H}$ is available for all channel realizations $q \in \mathcal{D}$. Although such information is not accessible in practical systems, this criterion is useful for analyzing the average behavior of the decomposition and identifying a representative rank for a given SNR level.

For each SNR value, a single common rank $R_{\mathrm{avg}}$ is selected for the entire dataset by minimizing the average reconstruction error:
\begin{equation}
R_{\mathrm{avg}}
=
\arg\min_{R \geq 1}
\frac{1}{|\mathcal{D}|}\sum_{q \in \mathcal{D}}
\left\| \mathcal{H}^{(q)} -
\sum_{r=1}^{R} \hat{\mathcal{H}}_{r}^{(q)} \right\|_F,
\end{equation}
where $\hat{\mathcal{H}}_{r}^{(q)}$ denotes the $r$-th extracted rank-1 component of the reconstructed channel tensor for the $q$-th realization. The constraint $R \geq 1$ is imposed to exclude the trivial zero-rank solution.

This strategy yields a single rank value shared across all channel realizations and UE antennas at a fixed SNR. To further analyze the reconstruction quality, we evaluate the error on each UE antenna slice:
\begin{equation}
E(\mathcal{X}_n, \hat{\mathcal{X}}_n) =
\| \mathcal{X}_n - \hat{\mathcal{X}}_n \|_F,
\label{eq:rel_err}
\end{equation}

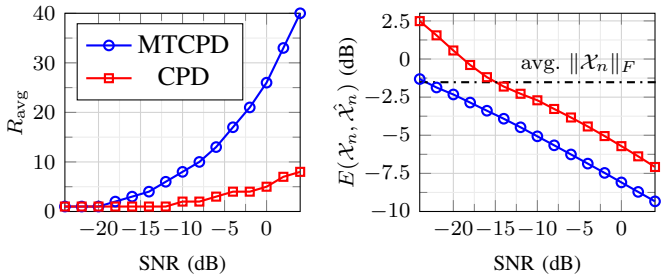
\begin{figure}[h]
\centering
\vspace{-2mm}
\begin{tikzpicture}

\begin{axis}[
    width=0.53\columnwidth,
    height=4.2cm,
    at={(0,0)},
    anchor=south west,
    xlabel={SNR (dB)},
    ylabel={$R_{\mathrm{avg}}$},
    ylabel style={yshift=-5pt},
    xmin=-24, xmax=4,
    ymin=0, ymax=40,
    xtick={-20,-15,-10,-5,0},
    ytick={0,10,20,30,40},
    minor x tick num=1,
    minor y tick num=1,
    grid=both,
    major grid style={gray!35},
    minor grid style={gray!15},
    tick label style={font=\footnotesize},
    label style={font=\footnotesize},
    legend style={at={(0.05,0.95)}, anchor=north west},
]

\addplot[
    blue,
    thick,
    mark=o,
    mark size=1.8pt
]
table[x=x, y expr={\thisrow{y3} == 0 ? 1 : \thisrow{y3}}, col sep=space] {rank_avg.dat};
\addlegendentry{MTCPD}

\addplot[
    red,
    thick,
    mark options={solid},
    mark=square,
    mark size=1.5pt
]
table[x=x, y expr={\thisrow{y4} == 0 ? 1 : \thisrow{y4}}, col sep=space] {rank_avg.dat};
\addlegendentry{CPD}

\end{axis}


\begin{axis}[
    width=0.53\columnwidth,
    height=4.2cm,
    at={(0.53\columnwidth,0)},
    anchor=south west,
    xlabel={SNR (dB)},
    ylabel={$E(\mathcal{X}_n,\hat{\mathcal{X}_n})$ (dB)},
    ylabel style={yshift=-5pt},
    xmin=-24, xmax=4,
    ymin=-10, ymax=3,
    xtick={-20,-15,-10,-5,0},
    ytick={-10,-7.5,-5,-2.5,0,2.5},
    minor x tick num=1,
    minor y tick num=1,
    grid=both,
    major grid style={gray!35},
    minor grid style={gray!15},
    tick label style={font=\footnotesize},
    label style={font=\footnotesize},
]

\addplot[
    blue,
    thick,
    mark=o,
    mark size=1.8pt
]
table[x=x, 
y expr=10*log10(\thisrow{y1}), 
col sep=space] {rank_avg.dat};

\addplot[
    red,
    thick,
    mark options={solid},
    mark=square,
    mark size=1.5pt
]
table[x=x,
y expr=10*log10(\thisrow{y2}),
col sep=space] {rank_avg.dat};

\addplot[
    black,
    thick,
    dash dot
]
coordinates {
    (-24, -1.521407)
    (8,   -1.521407)
};

\node[anchor=south east, font=\footnotesize]
at (3,-1.6)
{$\mathrm{avg.}\ \|\mathcal{X}_n\|_F$};

\end{axis}
\end{tikzpicture}
\vspace{-4mm}
\caption{Comparison of CPD and MTCPD as a function of SNR in terms of (left) the selected decomposition rank and (right) the corresponding reconstruction error per antenna slice.}
\label{fig:rank_nmse_vs_snr}
\end{figure}

Fig.~\ref{fig:rank_nmse_vs_snr} shows the selected dataset-averaged rank $R_{\mathrm{avg}}$ for different SNR levels (left), along with the corresponding minimum average reconstruction error $E(\mathcal{X}_n, \hat{\mathcal{X}}_n)$ per antenna slice (right) for both CPD and MTCPD.
It can be observed that the number of components required for channel estimation is generally higher for the proposed MTCPD method than for conventional CPD. This behavior is expected, as mode tensorization increases the effective separability of propagation components and enables the decomposition to resolve more detailed channel structures.

At the same time, despite using a larger number of components, MTCPD consistently achieves a lower reconstruction error than CPD across the entire SNR range. This indicates that the proposed MTCPD scheme provides a more accurate low-rank approximation of the channel tensor and leads to improved channel estimation performance.

\subsection{Phase-Coherence Metric for Steering Structure}

To quantify how closely a vector follows a steering (Vandermonde) structure, we introduce a metric based on phase increments.

Consider a vector $\mathbf{u} \in \mathbb{C}^{I}$, which may correspond to any canonical factor vector, e.g., $\hat{\mathbf{u}}_r^{(X)}$, $\hat{\mathbf{u}}_r^{(Y)}$, or $\hat{\mathbf{u}}_r^{(K)}$. Define the normalized phase ratios between adjacent elements as
\begin{equation}
\delta_i(\mathbf{u}) = \frac{\dot{u}_{i+1}}{\dot{u}_i}, \quad 
\dot{u}_i = \frac{u_i}{|u_i|}, \quad i=1,\ldots,I-1.
\end{equation}

For an ideal steering vector, these ratios are constant. Therefore, we define their sample mean
\begin{equation}
\bar{\delta}(\mathbf{u}) = \frac{1}{I-1} \sum_{i=1}^{I-1} \delta_i(\mathbf{u}),
\end{equation}
and quantify the deviation from the ideal structure using
\begin{equation}
\sigma_{\delta}(\mathbf{u}) = \sqrt{1 - \left|\bar{\delta}(\mathbf{u})\right|^2}.
\end{equation}

The metric $\sigma_{\delta}(\mathbf{u}) \in [0,1)$ measures the deviation from a steering structure: it approaches zero for vectors well approximated by a single plane wave and increases as the phase progression becomes less coherent.

In the proposed framework, this metric is evaluated for each factor vector $\hat{\mathbf{u}}_r^{(X)}$, $\hat{\mathbf{u}}_r^{(Y)}$, and $\hat{\mathbf{u}}_r^{(K)}$. To characterize the $r$-th component, we define an aggregated metric across the physical modes as
\begin{equation}
\sigma_r = \frac{1}{3} \left(
\sigma_{\delta}\big(\hat{\mathbf{u}}_r^{(X)}\big) +
\sigma_{\delta}\big(\hat{\mathbf{u}}_r^{(Y)}\big) +
\sigma_{\delta}\big(\hat{\mathbf{u}}_r^{(K)}\big)
\right).
\label{eq:phase_coherence_mean}
\end{equation}

Lower values of $\sigma_r$ indicate stronger consistency with the steering model and, consequently, a higher likelihood that the corresponding component represents a physical propagation path.

In the experiments, $\sigma_r$ is used as a component-selection criterion. Components with lower values of $\sigma_r$ are retained, while those with higher values are more likely to correspond to noise or non-structured contributions. The selection threshold is chosen empirically as $\sigma_r=0.5$ and is not further optimized in this work.

\subsection{Algorithm Performance in Terms of Spectral Efficiency}

Since the objective of channel estimation is to improve system-level performance, spectral efficiency is used as the target evaluation metric.
For each subcarrier $k=1,\ldots,K$, the estimated channel matrix $\hat{\mathbf{H}}_k \in \mathbb{C}^{M\times N}$ is used for precoder design. Let $P \leq \min(M,N)$ denote the number of transmitted streams. Assuming the singular value decomposition, $
\hat{\mathbf{H}}_k = \mathbf{U}_k \mathbf{S}_k \mathbf{V}_k^{\mathrm{H}},
$
the precoder is formed from the first $P$ left singular vectors of $\mathbf{U}_k$:
\begin{equation}
\hat{\mathbf{W}}_k =
\frac{1}{\sqrt{P}}
\left[
\mathbf{w}_{k,1},\ldots,\mathbf{w}_{k,P}
\right].
\end{equation}

The spectral efficiency for subcarrier $k$ is then computed using the true channel $\mathbf{H}_k$ as
\begin{equation}
C(\mathbf{H}_k,\hat{\mathbf{W}}_k)
=
\log \det
\left(
\mathbf{I}
+
\frac{1}{\sigma_{DL}^2}
\hat{\mathbf{W}}_k^{\mathrm{H}}
\mathbf{H}_k
\mathbf{H}_k^{\mathrm{H}}
\hat{\mathbf{W}}_k
\right),
\end{equation}
where $\sigma_{DL}^2$ is the downlink noise power. In simulations, $SNR_{\mathrm{DL}} = 10$ dB.

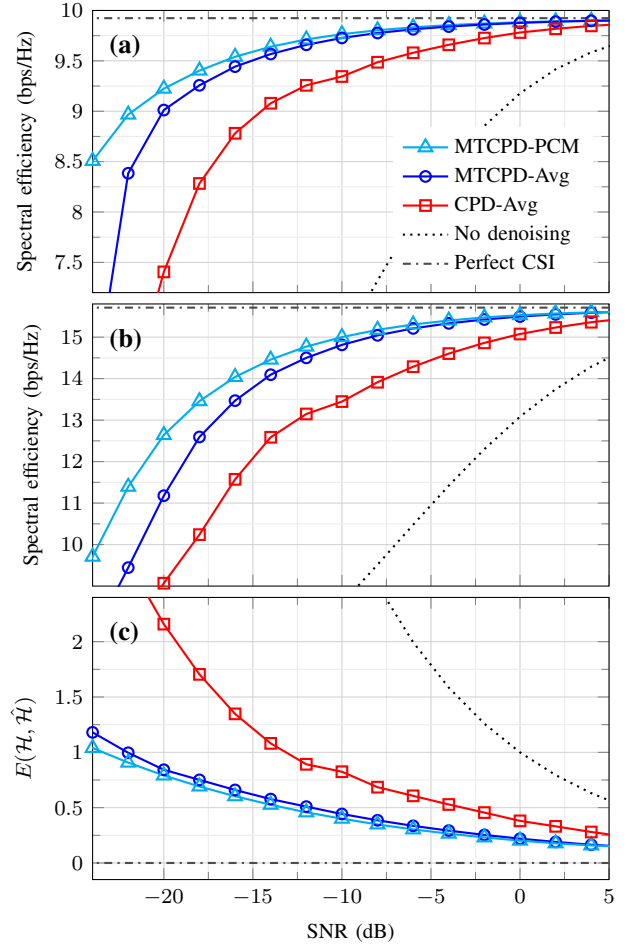
\begin{figure}[!t]
    \centering
    \begin{tikzpicture}

\pgfplotsset{
    finalResultsAxis/.style={
        width=0.95\linewidth,
        height=0.60\linewidth,
        xmin=-24,
        xmax=5,
        xtick={-20,-15,-10,-5,0,5},
        minor x tick num=1,
        minor y tick num=1,
        grid=both,
        major grid style={gray!35},
        minor grid style={gray!15},
        tick label style={font=\footnotesize},
        label style={font=\footnotesize},
        legend style={font=\footnotesize, draw=none},
        legend cell align={left},
    }
}

\begin{axis}[
    name=plotA,
    finalResultsAxis,
    ymin=7.2,
    ymax=10,
    ytick={7.5,8,8.5,9,9.5,10},
    ylabel={Spectral efficiency (bps/Hz)},
    xticklabel=\empty,
    xlabel={},
    legend pos=south east,
]

\addplot[cyan, thick, mark=triangle, mark size=3pt, mark options={solid}]
table[x=x, y=y2] {SU_rate_rank_1.dat};
\addlegendentry{MTCPD-PCM}

\addplot[blue, mark=o, thick, mark options={solid}]
table[x=x, y=y1] {SU_rate_rank_1.dat};
\addlegendentry{MTCPD-Avg}

\addplot[red, mark=square, thick, mark options={solid}]
table[x=x, y=y3] {SU_rate_rank_1.dat};
\addlegendentry{CPD-Avg}

\addplot[black, mark size=3pt, thick, dotted, mark options={solid}]
table[x=x, y=y4] {SU_rate_rank_1.dat};
\addlegendentry{No denoising}

\addplot[black!70, thick, dash dot]
coordinates {(-28, 9.923645) (8, 9.923645)};
\addlegendentry{Perfect CSI}

\node[anchor=north west, font=\normalsize\bfseries]
    at ([xshift=1mm,yshift=-2mm]current axis.north west) {(a)};

\end{axis}

\begin{axis}[
    name=plotB,
    finalResultsAxis,
    at={(plotA.south west)},
    anchor=north west,
    yshift=-1.5mm,
    ymin=9,
    ymax=15.8,
    ytick={10,11,12,13,14,15},
    ylabel={Spectral efficiency (bps/Hz)},
    xticklabel=\empty,
    xlabel={},
]

\addplot[blue, mark=o, thick, mark options={solid}]
table[x=x, y=y1] {SU_rate_rank_2.dat};

\addplot[cyan, thick, mark=triangle, mark size=3pt, mark options={solid}]
table[x=x, y=y2] {SU_rate_rank_2.dat};

\addplot[red, mark=square, thick, mark options={solid}]
table[x=x, y=y3] {SU_rate_rank_2.dat};

\addplot[black, mark size=3pt, thick, dotted, mark options={solid}]
table[x=x, y=y4] {SU_rate_rank_2.dat};

\addplot[black!70, thick, dash dot]
coordinates {(-28, 15.711467) (8, 15.711467)};

\node[anchor=north west, font=\normalsize\bfseries]
    at ([xshift=1mm,yshift=-2mm]current axis.north west) {(b)};

\end{axis}

\begin{axis}[
    name=plotC,
    finalResultsAxis,
    at={(plotB.south west)},
    anchor=north west,
    yshift=-1.5mm,
    ymin=-0.15,
    ymax=2.4,
    ytick={0,0.5,1,1.5,2},
    xlabel={SNR (dB)},
    ylabel={$E(\mathcal{H}, \hat{\mathcal{H}})$},
]

\addplot[blue, mark=o, thick, mark options={solid}]
table[x=x, y=y1] {Rec_err.dat};

\addplot[cyan, thick, mark=triangle, mark size=3pt, mark options={solid}]
table[x=x, y=y2] {Rec_err.dat};

\addplot[red, mark=square, thick, mark options={solid}]
table[x=x, y=y3] {Rec_err.dat};

\addplot[black, mark size=3pt, thick, dotted, mark options={solid}]
table[x=x, y=y4] {Rec_err.dat};

\addplot[black!70, thick, dash dot]
coordinates {(-28, 0) (8, 0)};

\node[anchor=north west, font=\normalsize\bfseries]
    at ([xshift=1mm,yshift=-2mm]current axis.north west) {(c)};

\end{axis}

\end{tikzpicture}
    \caption{Median performance comparison versus SNR: (a) rank-1 SU-MIMO spectral efficiency, (b) rank-2 SU-MIMO spectral efficiency, and (c) reconstruction error.}
    \label{fig:final_results}
\end{figure}

Figure~\ref{fig:final_results}(a,b) shows the median spectral efficiency as a function of SNR for rank-1 and rank-2 transmission regimes, respectively. Fig.~\ref{fig:final_results}(c) presents the corresponding reconstruction error \eqref{eq:rel_err} evaluated on the full channel tensors.

The labels MTCPD-Avg and MTCPD-PCM denote the proposed MTCPD method with dataset-averaged rank selection and the phase-coherence-based criterion, respectively, while CPD-Avg corresponds to the conventional CPD approach with dataset-averaged rank selection.

The results demonstrate that both MTCPD-based schemes consistently outperform the conventional CPD-based channel estimation across the entire SNR range. The performance gap becomes more pronounced at low SNR values, where noise has a stronger impact and CPD is more likely to capture noise components rather than meaningful channel structures.

For example, at $\mathrm{SNR}_{UL} = -20$~dB, MTCPD-Avg increases the median spectral efficiency by approximately 1.6~bps/Hz and 1.8~bps/Hz compared to CPD-Avg for rank-1 and rank-2 transmission, respectively. The MTCPD-PCM scheme achieves even larger gains of about 1.8~bps/Hz and 3.5~bps/Hz for rank-1 and rank-2 transmission, respectively.

Among the considered rank selection strategies, the proposed phase-coherence-based criterion provides the best overall performance. This can be attributed to the metric $\sigma_r$, which suppresses components that do not exhibit a Vandermonde structure and are therefore less likely to correspond to physical propagation paths. As a result, the selected components better preserve the underlying channel structure, leading to more accurate channel reconstruction and improved precoder design.

The observed gains in spectral efficiency are consistent with the lower reconstruction error shown in Fig.~\ref{fig:final_results}(c), confirming that improved tensor approximation directly translates into better system-level performance.

\section{Conclusion}
This paper proposed a mode-tensorized canonical polyadic decomposition (MTCPD) method for channel estimation in MIMO systems. By exploiting the Vandermonde structure of array and frequency response vectors through tensor reshaping into a higher-order representation, the approach improves the separability of propagation components and the effectiveness of CP decomposition.
Numerical results demonstrate that the proposed MTCPD method outperforms conventional CPD and improves robustness to noise, particularly for low-SNR conditions. The use of a phase coherence metric (PCM) further enhances performance by selecting informative rank-1 components and suppressing those that do not correspond to physically meaningful paths. The combined MTCPD framework with PCM-based rank selection achieves the best performance over the SNR range from $-24$ dB to $4$ dB, with a median spectral efficiency gain of approximately $3.5$ bps/Hz compared to conventional CPD for rank-2 transmission.

\end{document}